\documentclass[12pt,a4paper,english]{article}

\usepackage{float}
\usepackage[slantedGreek]{mathpazo}
\usepackage[pdftex]{graphicx}
\usepackage{amsfonts}
\usepackage{setspace}
\usepackage{amssymb}
\usepackage{amsthm}
\usepackage{graphicx}
\usepackage{amsmath,scalerel}%
\usepackage{subcaption}
\usepackage{color}
\usepackage[english]{babel}
 
\usepackage{natbib}
\usepackage{graphicx}
\usepackage{amsmath, amsthm, amssymb, amsfonts}
\usepackage{bm}
\usepackage{epstopdf}
\usepackage{times}
\usepackage{hyperref}
\usepackage{booktabs}
\usepackage{array}
\usepackage{lscape}
\usepackage{tikz}
\usepackage{algorithm2e}
\usepackage{caption}
\usepackage[all]{hypcap}                    
\usepackage{babel}

\usepackage{epstopdf}

\theoremstyle{plain}

\begin{document}

	\newpage
	
	\begin{center}
		\Large
		\textbf{The Illusion of the Illusion of Sparsity:}\\ 
		\textbf{An exercise in prior sensitivity}
		
		\vspace{0.4cm}
		\large
		Bruno Fava, Northwestern University, USA\\
		and\\
		Hedibert F. Lopes, Insper, Brazil
		
		\vspace{0.9cm}
		
		This draft: September 9th 2020.
		
		\vspace{0.9cm}
		
		\textbf{Abstract}
		
	\end{center}
	
	\noindent
	The emergence of Big Data raises the question of how to model economic relations when there is a large number of possible explanatory variables. We revisit the issue by comparing the possibility of using dense or sparse models in a Bayesian approach, allowing for variable selection and shrinkage. More specifically, we discuss the results reached by \cite{glp} through a ``€œSpike-and-Slab''€ prior, which suggest an ``€œillusion of sparsity''€ in economic data, as no clear patterns of sparsity could be detected. We make a further revision of the posterior distributions of the model, and propose three experiments to evaluate the robustness of the adopted prior distribution. We find that the pattern of sparsity is sensitive to the prior distribution of the regression coefficients, and present evidence that the model indirectly induces variable selection and shrinkage, which suggests that the ``€œillusion of sparsity''€ could be, itself, an illusion.  Code is available on Github\footnote{github.com/bfava/IllusionOfIllusion}.\\
	
	\noindent
	Keywords: Sparsity, Model selection, High Dimensional Data, Shrinkage, Bayesian Econometrics.


	
	
	
	\section{Introduction}
	
    It''€™s the Big Data Era. While Tech Giants revolution the markets of the U.S. and China, economists are still adapting to the new flow of data and studying how to incorporate them in the research agenda. In the presence of many data sources, it''€™s a common situation to have a large number of variables that can possibly determine a variable of interest, so that the number of regressors approximates or exceeds the number of observations itself. This article addresses the issue of how to deal with the situation, considering whether it is wiser to use all the available regressors or using methods that define which are the most important ones.
    
    In datasets in which the regressors outnumber the observations, the use of classical estimation methods, such as the Ordinary Least Squares (OLS), is not even possible as the statistical inference would be based on a negative number of degrees of freedom. Even when there is a small positive number of degrees of freedom, the OLS estimator drives very poor results, once it''€™s expected overfitting and high degrees of multicollinearity. Many methods have been developed to deal with the problem, using classical and Bayesian statistics, and Machine Learning (ML) schemes, such as Random Forests (RF) that consider nonlinearity in the mean function and can deal with a large number of variables.
    
    Even though in the literature it has been identified some classes of models that perform well for predictions, not everything in Economics is about prediction. A vast class of articles in Economics focus on the individual impact of some key regressors on a response variable and, therefore, models like the RF may not be adequate, as they make it difficult to interpret the individual effects of each regressor, in addition to driving biased estimates of partial effects.
    
    It is then convenient to look at the so-called sparse models, that in the presence of many predictors, select the most important ones. The counterpart are the dense models, that instead of choosing some variables despite the others, consider all of them, shrinking the estimated coefficients towards a zero mean so that, despite the relatively small sample size, overfitting is avoided.
    
    A series of models were developed that consider sparsity in explanatory variables, for example, the famous Least Absolute Shrinkage and Selection Operator, the LASSO, introduced by \cite{tibshirani1996regression}. By defining a constant limit for the sum of the absolute value of the coefficients in a model, the LASSO shrinks the coefficients towards zero and, by doing so, estimates some of them to be exactly zero, that is, excludes those variables from the model. This kind of design does solve the problem of the big number of predictors by using statistical inference to determine the ones that are the most important, and then allowing for an easy interpretation of partial effects.
    
    Still, the choice to use sparse models may not be a free lunch. A recent work developed by \cite{glp} ``€" henceforth referred to as GLP ``€" explored the suitability of sparse modeling for economics series. They took two datasets in Microeconomics, Macroeconomics and Finance, and defined a ``€œSpike-and-Slab''€ prior for the coefficients of linear predictive models, following \cite{mitchell1988bayesian}. This prior was chosen because, by taking a probability $q$ of inclusion of each predictor as an unknown parameter with a uniform prior, it allows the model to take both the sparse or dense designs, hence not assuming one of them, and making inference on which of the possibilities is more probable.
    
    The results are not encouraging for those who prefer adopting sparse representations: the model drove a not-sparse design for five of the six applications, baptizing the title of their article as ``€œEconomic predictions with big data: The illusion of sparsity''€. The authors conclude that sparsity should not be simply assumed when modeling an economics series, as that is uncertain, and therefore should only be used in the presence of strong statistical evidence.
    
    This work proposes a revision of the methods adopted by GLP. We reproduce the model they used, a ``€œSpike-and-Slab''€ prior distribution, that considers, in a linear model, a probability $q$ of inclusion of each predictor, while the included coefficients are modeled as draws from a Gaussian distribution. The variance from this distribution is defined as $\gamma^2$, that thus controls the degree of shrinkage. By treating both hyperparameters as random variables, they conducted Bayesian inference on them to visualize whether there would be greater concentration in small values of $q$ or a more important dependence on greater shrinkage, that is, if the dataset should be treated mainly as sparse or dense.
    
    We use five of the six original datasets from GLP (the Micro 1 and 2, Macro 1 and 2, and Finance 1 datasets) and reproduce the algorithm for estimating the model, first reinterpreting the posterior distributions, and then proposing three experiments to evaluate how well the model behaves in controlled environments. 
    
    First, we analyze the posterior distribution of the coefficients of the linear model, when included, what was not explored in GLP. It indicates a certain inability of the model in distinguishing whether a variable should be excluded, or included with a very small coefficient, what would result in the overestimation of the probability of inclusion, and could help explain the results achieved. Second, we add completely random variables as possible predictors to the datasets, and find that the model is able to correctly exclude them only in a sub-selection of the datasets.
    
    Third, it is proposed a modification to the prior distribution of the parameters of the linear model, by fitting a t-student distribution instead of a Gaussian, allowing for fatter tails. The heavier-tailed distribution was more restrictive in selecting possible predictors, and results once again corroborate with the thesis that the original Spike-and-Slab prior is unable to correctly allow and distinguish between shrinkage or sparsity. Finally, it is developed a simulation study to check the performance of the original model and with the t-student modification in a totally controlled environment. At the same time that both approaches don't present great performance, the analysis of the posterior distributions reinforces the belief that the adopted prior distribution incorrectly induces shrinkage. 
    
    All the evidence raised allows this paper to conclude that the Spike-and-Slab approach does not seem robust, and could lead to the illusion that sparsity is nonexistent, when it might exist.
    
    The rest of the article is organized as follows.  In Section 2 we explore the article from GLP, explain thoroughly the model used, and discuss the main results found in the paper. In Section 3 we propose the three experiments: adding random variables to the datasets, modifying the prior distribution of the coefficients from a normal to a t-student distribution, and finally a simulation study. In Section 4, we present a conclusion.

    \section{Revisiting GLP}

    In this section, we reproduce and explore the analysis made by \cite{glp}, with a ``€œSpike-and-Slab''€ prior distribution for a linear predictive model applied in different economics-related datasets. GLP selected six popular datasets they consider ``€œbig data''€, for the relatively large number of predictors compared to the number of observations: two in Macroeconomics, two in Microeconomics and two in Finance. From the six settings, we don't consider only the Finance 2 dataset.  Nonetheless, we briefly review the existing literature on Bayesian sparsity that motivated \cite{glp} findings.
    
\subsection{A brief review of Bayesian regularization}
For the sake of space, let us consider the standard Gaussian linear model, already in a matrix format,
$$
y = X\beta + \epsilon, \qquad \epsilon \sim N(0,\sigma^2I_n),
$$
and RSS$=(y-X\beta)'(y-X\beta)$ be the residual sum of squares.  Two of the most popular forms of regularization arise from a {$\ell_2$-penalty} ridge regression of \cite{hoerl1970ridge} or a {$\ell_1$ penalty} lasso regression of \cite{tibshirani1996regression}:
\begin{eqnarray*}
\hat{\beta}_{ridge}&=& \mbox{arg} \min_\beta \left\{RSS +\lambda_r^2 \sum_{j=1}^q \beta_j^2\right\},  \qquad \lambda_r^2 \geq 0,\\
&=&(X'X+\lambda_r^2 I_q)^{-1}X'y\\
\hat{\beta}_{lasso} &=& \mbox{arg} \min_\beta \left\{RSS +\lambda_l \sum_{j=1}^q |\beta_j| \right\},  \qquad \lambda_l \geq 0,
\end{eqnarray*}
which can be solved by a \textit{coordinate gradient descent} algorithm.

As it is well established, both ridge and lasso estimates are essentially posterior modes.  Broadly speaking, the posterior mode, or the maximum a posteriori (MAP), estimate is given by
$$
{\tilde \beta}_{\mbox{mode}} = \mbox{arg} \min_\beta \{-2\log p(y|\beta) - 2\log p(\beta)\}.
$$
The $\hat{\beta}_{ridge}$ estimate, hence, equals the posterior mode of the normal linear model with $p(\beta_j) \propto \exp\{-0.5\lambda_r^2 \beta_j^2\}$, which is a Gaussian distribution with location 0 and scale $1/\lambda_r^2$, $N(0,1/\lambda_r^2)$.
The mean is $0$, the variance is $1/\lambda_r^2$ and the excess kurtosis is {\color {red} $0$}.  Similarly, the $\hat{\beta}_{lasso}$ estimate equals the posterior mode of the normal linear model with
$p(\beta_j) \propto \exp\{-0.5\lambda_l|\beta_j|\}$, which is a Laplace distribution with location 0 and scale $2/\lambda_l$, Laplace$(0,2/\lambda_l)$.
The mean is $0$, the variance is $8/\lambda_l^2$ and excess kurtosis is $3$.

As a matter of fact, a whole family of regularization schemes arise, as suggestd by \cite{ishwaran2005spike}, by defining spike and slab model as a Bayesian model specified by the following prior hierarchy:
\begin{eqnarray*}
(y_t|x_t,\beta,\sigma^2) &\sim& N(x_t' \beta,\sigma^2), \qquad t=1,\ldots,n\\
(\beta|\psi) &\sim& N(0,\mbox{diag}(\psi))\\
\psi &\sim& \pi(d \psi)\\
\sigma^2 &\sim& \mu(d \sigma^2).
\end{eqnarray*}
The distribution chosen to model $\psi$ defines what kind of shrinkage and selection strategy is being used.

Alternative choices for $\psi$ appear, amongst many others, in the
two-component spike-and-slab-type prior of \cite{george1993variable},
the Laplace prior of \cite{park2008bayesian}, 
Normal-Gamma prior of \cite{griffin2010}, the horseshoe prior of
Carvalho, Polson and Scott (2010) and the Dirichlet''€"Laplace prior of 
\cite{dunson2015} and the spike-and-slab lasso of \cite{rockova2018}.  See also \cite{hahn2019} who introduced an efficient sampling scheme for Gaussian linear regression with arbitrary priors.

\subsection{The Model}
    
    Given a response variable $y_t$, a vector of possible predictors $x_t$, of size $k$, and a vector of always-included variables $u_t$, of size $l$, with generally $k \gg l$, the model is defined as:
    
    $$y_t = u_{t}^{'}\phi + x_{t}^{'}\beta + \epsilon_t$$
    
    Where $\epsilon_t$ is an i.i.d. stochastic Gaussian error term with zero mean and variance $\sigma^2$. For simplification, all variables included are standardized to have zero mean and variance one. The vector $\phi$  will never contain zeros, as the predictor included in $u_t$ are always taken as relevant to the regression. The vector $\beta$, otherwise, is supposed to inform the suitability of whether a dense or sparse representation. Thus, most of the elements of this vector may be zero ``€" defining a sparse model ``€", or non-zero ``€" a dense model. To reflect the possibility of taking one of both representations, the following prior distribution is proposed for the unknown parameters $\sigma^2,\phi,\beta$:

    $$p(\sigma^2) \propto \frac{1}{\sigma^2}$$
    $$\phi \propto \text{flat}$$
    $$\beta_i | \sigma^2, \gamma^2, q \sim \left\{\begin{matrix}
    
    N(0,\sigma^2 \gamma^2 ) & \text{with prob. }q \\
    
    0 & \text{with prob. } 1 - q 
    
    \end{matrix} \ \ \ i = 1, ..., k.\right.$$

	Where the prior for the variance $\sigma^2$ is the improper Jeffrey''€™s prior, the prior for $\phi$ is uninformative, and each of the parameters in the vector $\beta$ can be either zero, with probability $1-q$, or a draw from a Gaussian distribution with zero mean and variance $\sigma^2 \gamma^2$, with probability $q$. The hyperparameter $\gamma^2$ controls the degree of shrinkage: the larger $\gamma^2$, the smaller the shrinkage, as the regression coefficients can be more distant from zero.
	
	The prior distribution for the hyperparameter $\gamma^2$ is induced by a prior on a transformation of the coefficient. Specifically, they set a prior for the coefficient of determination, $R^2$:
	
	\begin{center}
	   
	   $R^2(\gamma^2, q) \equiv \dfrac{qk\gamma^2\overline{v}_x}{qk\gamma^2\overline{v}_x + 1}$
	    
	\end{center}
	
    Where $\overline{v}_x$ is the sample average variance of the predictors. The prior distribution of the hyperparameters is then defined by uniform distributions:

    \begin{center}
        
        $q \sim Beta(1, 1)$
        
        $R^2 \sim Beta(1, 1)$
        
    \end{center}

    \subsection{Inclusion Probabilities}
    
    \begin{figure}[t]
    \centering
        \includegraphics{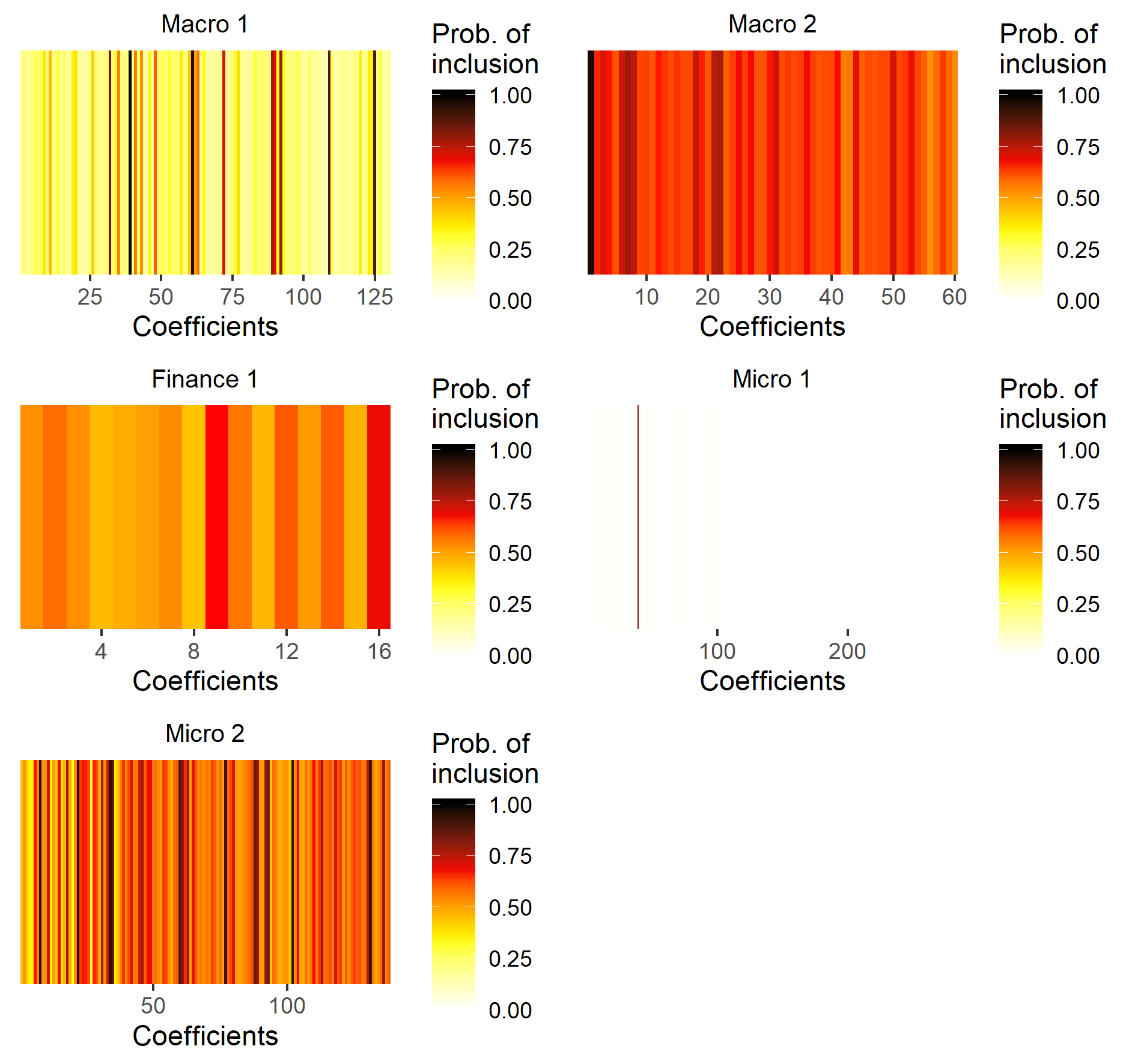}
        \caption{Probability of inclusion of each predictor}
        \label{fig:vars}
    \end{figure}
    
    The heatmaps in figure \ref{fig:vars} show the posterior probability of inclusion of each predictor in the model, that is, the percentage of times that each covariate was included on the Markov Chain Monte Carlo estimation of the model. Thus, for example, if a stripe presents a near-black color, it indicates that such predictor was included on nearly all of the iterations of the estimation, that is, its probability of inclusion is close to 100\%. On the other hand, if the stripe color is light yellow, its probability of inclusion is small.
    
    \subsection{GLP Conclusion}
    
	After analyzing the posterior distributions, GLP investigate whether a pattern of sparsity can be identified in the datasets, by measuring the percentage of times each variable was included in the regression (figure \ref{fig:vars}). The conclusion is that a clear pattern of sparsity is found only on the Micro 1 dataset, in which only one variable is included most of times. For all other datasets it''€™s not possible to distinguish which variables should be included, as many have a high estimated probability of inclusion. That indicated that a dense model, that allows for the selection of many variables while shrinking their coefficients, should be the most adequate for them.
	
	Thus, even when the estimated number of included variables is small, it might not be easy to determine what the pattern of sparsity should be, that is, which variables should be selected. This result allows GLP to conclude that sparsity cannot be assumed for any economic dataset, unless in the presence of strong statistical evidence, and suggest an "illusion of sparsity" when using statistical models that assume (and force) sparsity.
   
   \subsection{Posterior distribution of $\beta_i | (\beta_i \neq 0)$}
    
\begin{figure}[t]
        \centering
        \includegraphics[height=10cm, width=17cm]{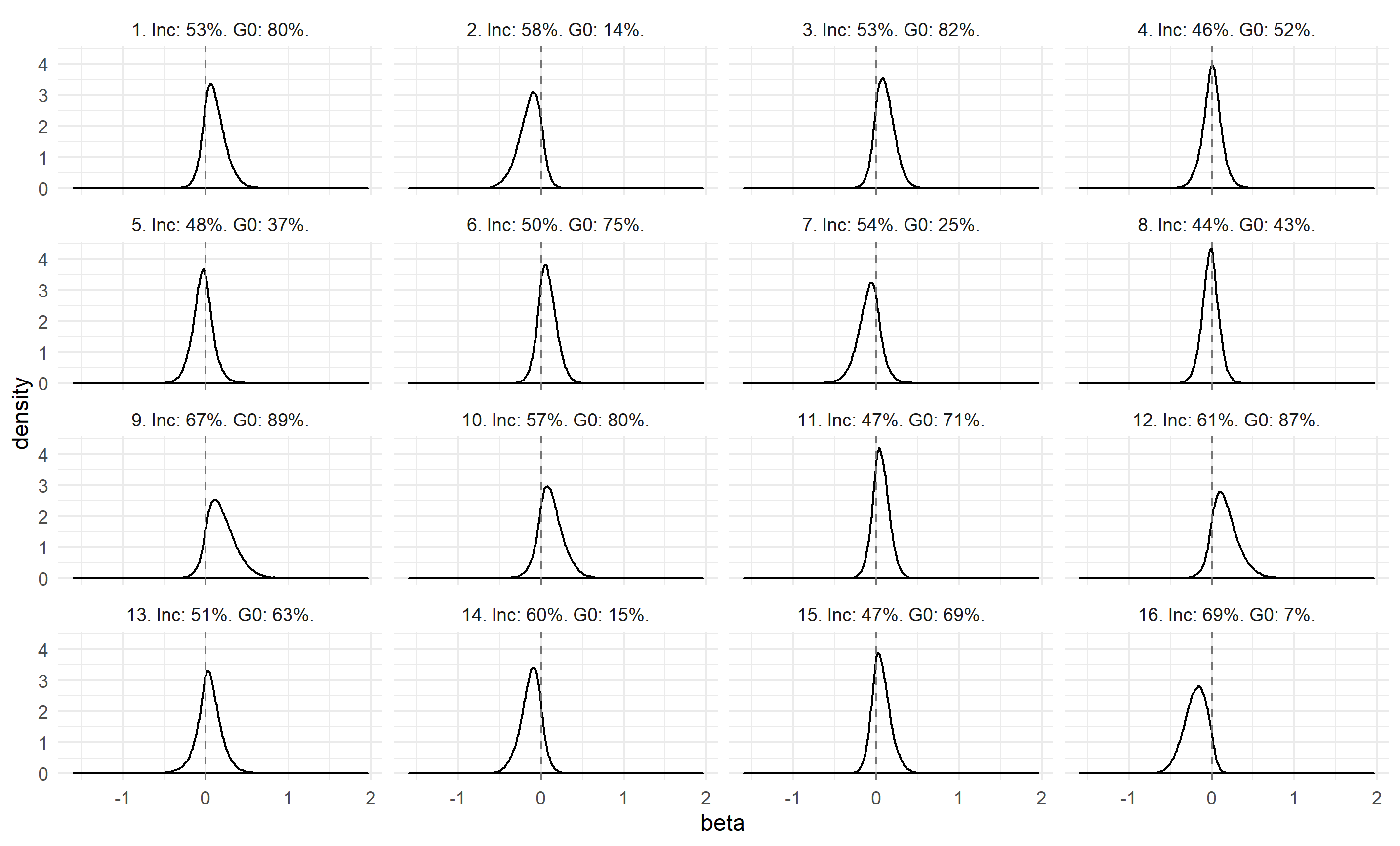}
        \caption{Posterior distribution of beta for the Finance 1 dataset. "Inc." means the probability of inclusion and "G0" the probability of being greater than zero}
        \label{fig:beta_fin1}
        \text{}
    \end{figure}

\begin{figure}[t]
        \centering
        \includegraphics[height=10cm, width=17cm]{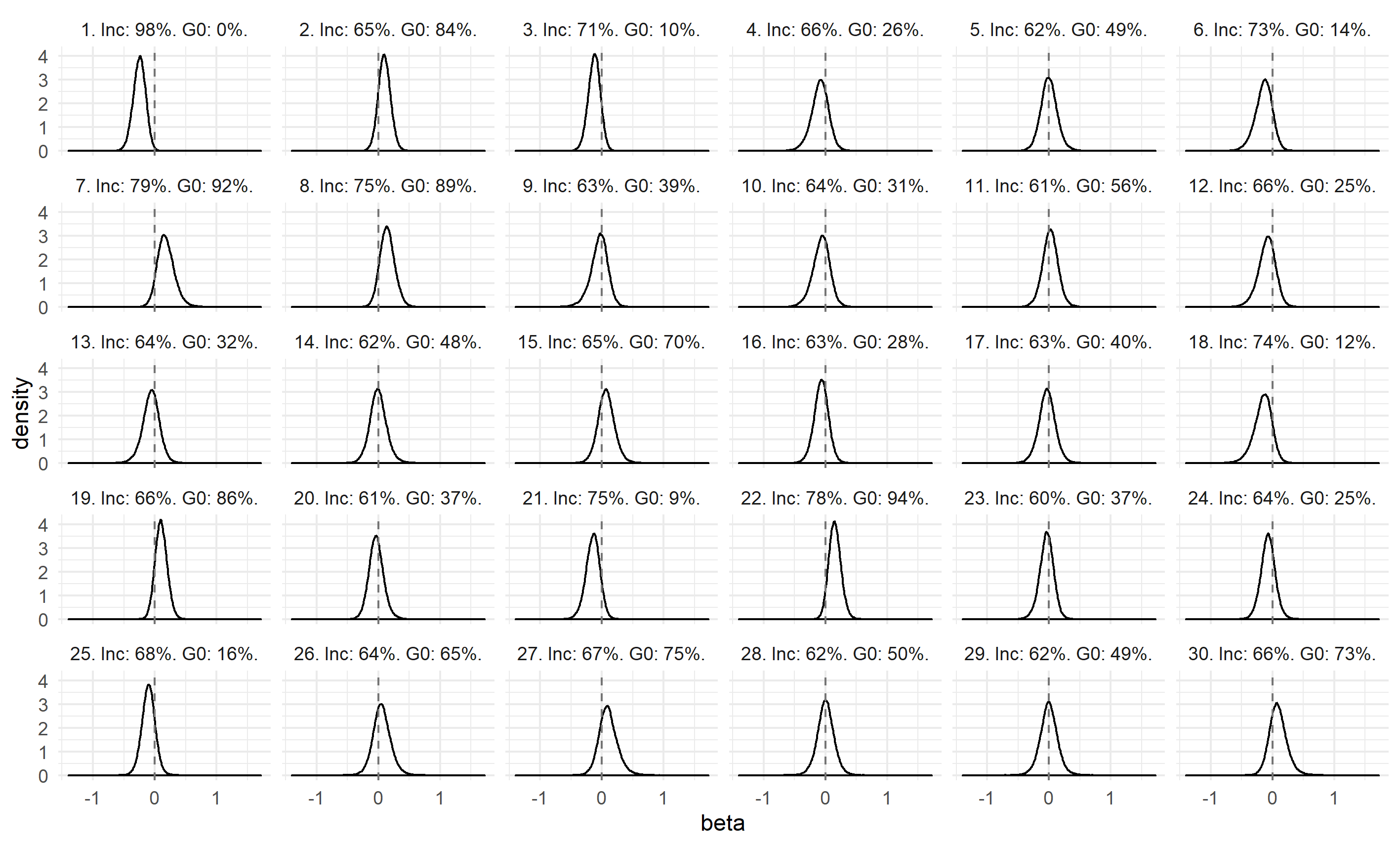}
        \caption{Posterior distribution of beta for the Macro 2 dataset (1/2). "Inc." means the probability of inclusion and "G0" the probability of being greater than zero}
        \label{fig:beta_macro2_1}
    \end{figure}

\begin{figure}[t]
        \centering
        \includegraphics[height=10cm, width=17cm]{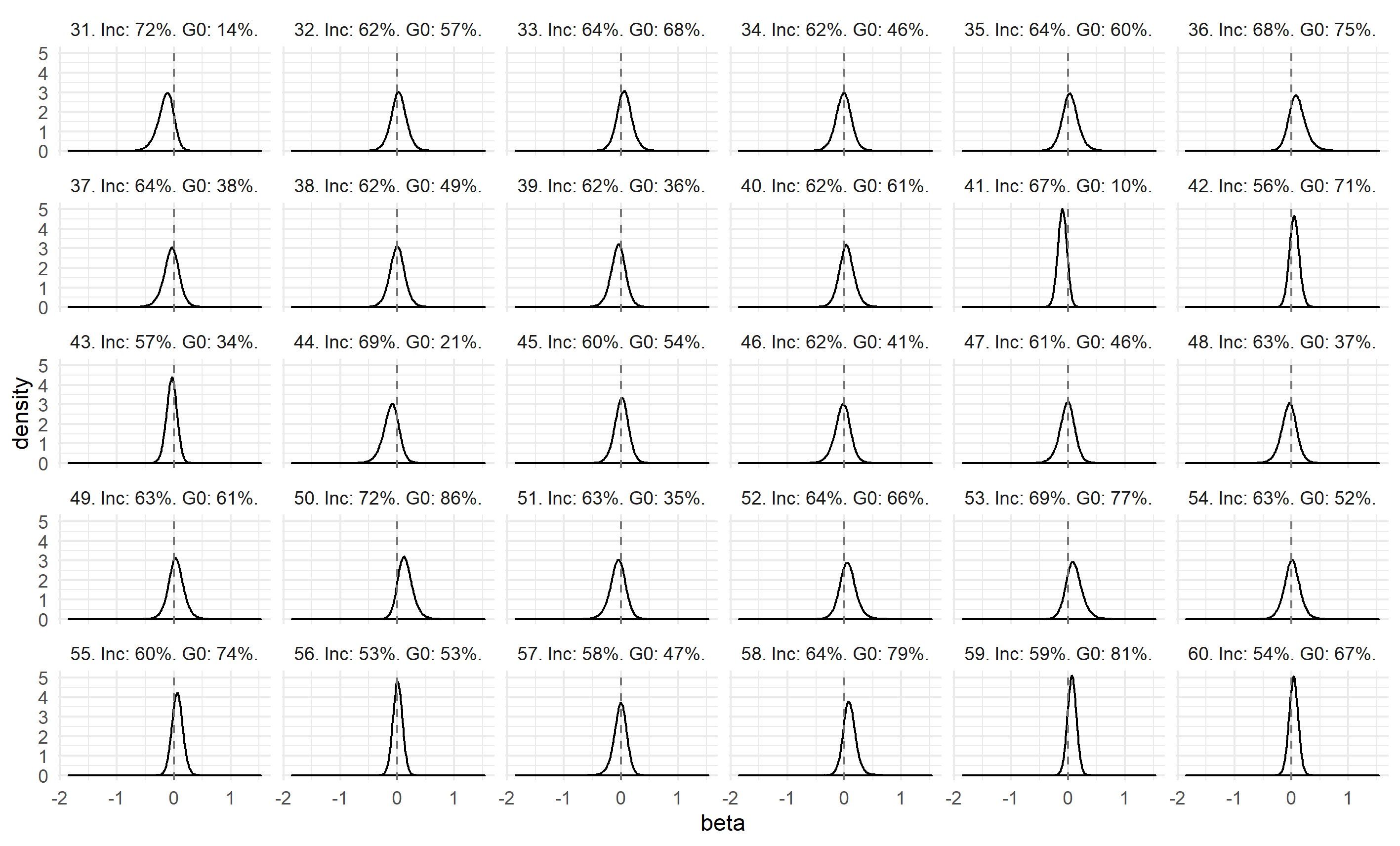}
        \caption{Posterior distribution of beta for the Macro 2 dataset (2/2). "Inc." means the probability of inclusion and "G0" the probability of being greater than zero}
        \label{fig:beta_macro2_2}
    \end{figure}

    We present the posterior distribution of the coefficients $\beta_i | (\beta_i \neq 0)$ for all possible predictors, which was not presented in GLP. We focus on the Finance 1 and Macro 2 datasets only, because of the convenience that they present a smaller number of covariates - 16 and 60, respectively. The posterior for Macro 2 is divided in figures \ref{fig:beta_macro2_1} and \ref{fig:beta_macro2_2}, each with 30 predictors. The distributions for the Finance 1 are shown in figure \ref{fig:beta_fin1}.
    
    We evaluate how significant an included predictor is by analyzing how close to zero the coefficients of the included variables are. In order to clarify the results, on the graphics we show first the number (index) of the predictor followed by the number of "Inc", the probability of inclusion of that predictor (the same number plotted on figure \ref{fig:vars}), and the number of "G0", the probability that the predictor is greater than zero, that is, the percentage of the times that the estimated coefficient was positive, considering only the cases when the variable was included in the model.
    
    At the same time that it is expected that included coefficients be shrunken towards zero - specifically in a greater rate as the probability of inclusion grows -, if the concentration of the posterior distribution around zero is too large, it can be argued that the model can be failing in determining whether a predictor is relevant for fitting the model or not. That is, if the distribution of $\beta_i | (\beta_i \neq 0)$ is very concentrated around zero, the likelihood that $z_i = 1$ or $z_i = 0$ will be very close, as the inclusion of the coefficient would have very small impact on the regression.
    
    Therefore, the probability of inclusion $q$ might be overestimated, and some of the coefficients included on figure \ref{fig:vars} with high probability may be performing an almost negligible role in the model, with the explanatory capacity of the covariates concentrated on a few predictors. That is, a pattern of sparsity might be hidden on the many selected variables, what would imply that the prior distributions set are themselves inducing "density" and shrinkage, despite the goal of learning statistically whether shrinkage or selection is ideal.
    
    The graphics in figures \ref{fig:beta_fin1} to \ref{fig:beta_macro2_2}, reveal some interesting features of the posterior distribution. First, in fact, some predictors included have very concentrated and symmetrical posterior distributions around zero, indicating that if an economist were to define a pattern of predictors to include in a linear model, from the learning of the posterior distribution of $\beta$ they would very probably exclude these covariates, even when the plot in figure \ref{fig:vars} would indicate the opposite.
    
    This is very clearly the case, for example, in the Finance 1 setting in figure \ref{fig:beta_fin1}, of the variables 4 and 8, that are very concentrated and symmetrical around zero, with nearly half of the distribution to each of the sides of zero, despite having a probability of inclusion of 46\% and 44\%, respectively. Other variables, on the other hand, do present a very distinct pattern when included. For example, predictors 2, 3, 9, 12 and 16 have a large concentration of their coefficients away from zero, at the same time that their probability of inclusion is not largely superior to the ones of predictors 4 and 8. Predictor 3, for example, is included only 53\% of the times, and predictor 2 58\%.
    
    Other covariates show a more peculiar and dubious behavior. Predictor 11, for instance, has its coefficient clearly more concentrated to positive values, at the same time that it is very concentrated around zero, and is included only 47\% of times. It is hard to conclude from figure \ref{fig:beta_fin1} that a clear pattern of sparsity can be distinguished. Still, it seems clear the greater importance of a few variables in spite of others, as is the case of predictors 2, 9, 12, 14 and 16, and a smaller importance of other, such as 4, 5, 6, 8, 11, 13 and 15. Even though there seems to exist a direct relation between these patterns and the probability of inclusion - the least included predictor among the "most important" was included 58\% of times, whereas the most included among the "least important" was included 51\% of times -, interpreting sparsity from the graphic on figure \ref{fig:vars} by itself is misleading, and hides some important information behind each coefficient.
    
    As for the posterior of $\beta$ for the Macro 2 dataset, similar conclusions can be drawn. While some predictors are undoubtedly important for the model, such as predictor 1 (included 98\% of times always with a negative coefficient), others have a coefficient very close to zero when included in the model, such as predictors 5, 28, 29, 32, 34, 38, 54 and 57. Still, at the same time that one economist could easily exclude such predictors from a regression model, some of them are included in the spike-and-slab with a significant probability, of at least 60\%. It is also interesting to notice that, for example, predictor 12 is highly offset from zero, with 75\% of the distribution on negative values, but also presents a relatively small percentage of inclusion, of 66\%, while predictor 28 is highly symmetric around zero, with 50\% of the distribution in positive numbers, and is included almost at the same rate, 62\% of the times.
    
    This analysis let us conclude that even though, in fact, a distinct pattern of sparsity cannot be identified on the datasets, the spike-and-slab prior, as defined, seems to be itself inducing density and shrinkage, by including frequently many predictors with a near-zero coefficient.
    
    \section{Experiments}

    This section is composed of three parts. We first explore the power of selection of the spike-and-slab prior as specified, by adding random variables as additional predictors in the five datasets, and checking whether the posterior distribution was able to correctly identify their exclusion. Second, this article proposes a change in the model, by substituting the Gaussian prior distribution of the coefficient of the possible predictors for a t-student distribution. Finally, we develop a simulation study to check the conditions under which the model correctly selects a sparse model.
    
    The estimation algorithms were reproduced in R with Rcpp (R C++), and all the estimation code used in this whole session is displayed on Github\footnote{github.com/bfava/IllusionOfIllusion}. 
    
    \subsection{Adding Random Variables}
    
    \begin{figure}[t]
    \centering 
        \includegraphics{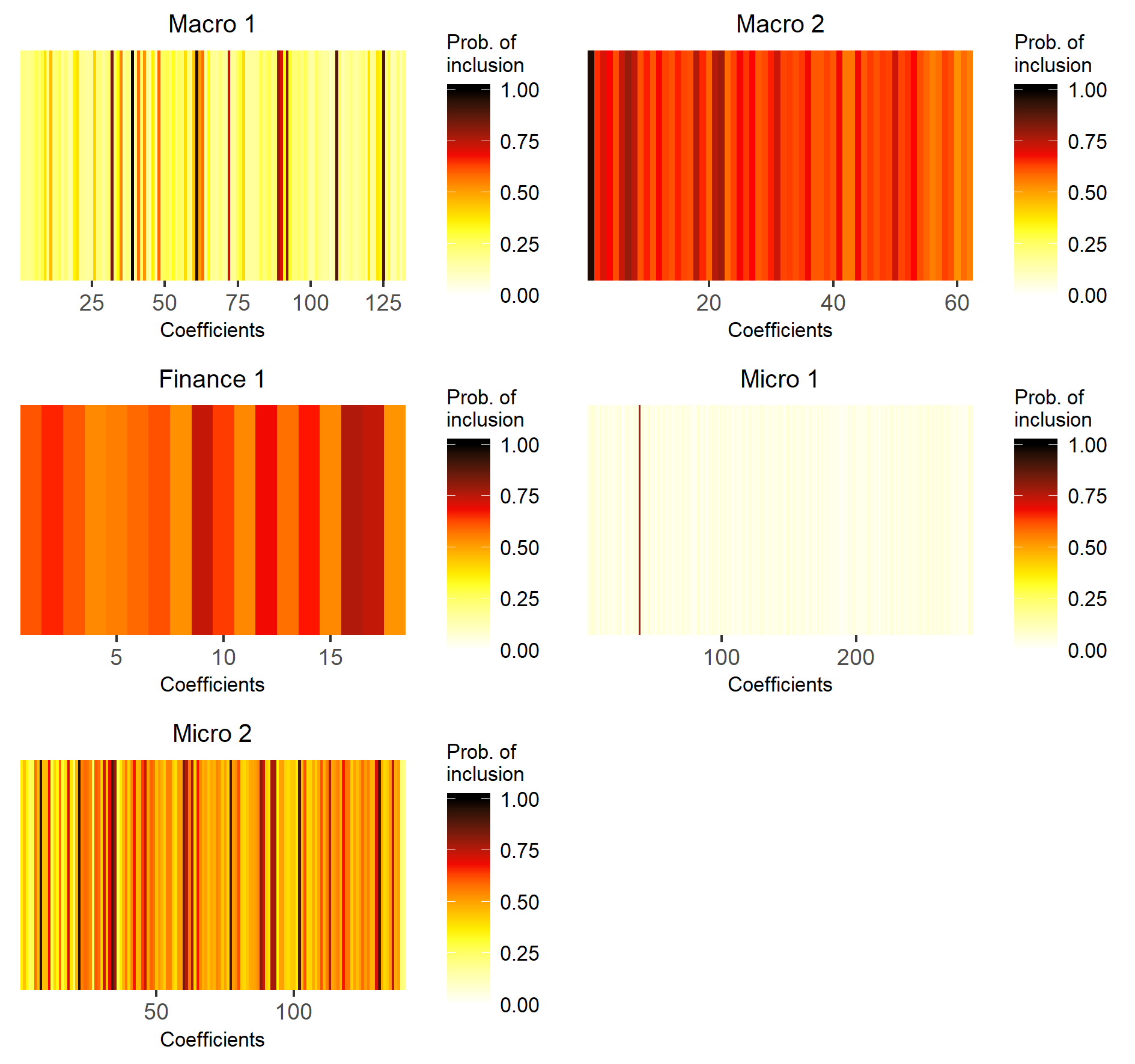}
        \caption{Probability of inclusion of each predictor - two last stripes are random variables}
        \label{fig:rnd_vars}
    \end{figure}
    
    In order to further explore the thesis proposed on the last topic -- that is, that the Spike-and-Slab prior might be itself inducing density --, we now propose a further experiment. We re-run the estimation algorithm for all the five datasets but now include two additional regressors that were completely randomly generated from a normal distribution, and re-scaled to have zero mean and standard deviation of one, like all the other predictors.
    
    The goal is to analyze if the model is able to determine the exclusion of such covariates, that are known to have no predictive power and fitting to the data, having only possibly spurious correlations.
    
    The graphic in figure \ref{fig:rnd_vars} brings the probability of inclusion of each predictor, where the last two stripes are the randomly generated predictors. This graphic is just like the one of figure \ref{fig:vars}, re-estimated with the new variables. It is easy to notice that the pattern of inclusion of the original predictors is very similar between the two figures.
    
    The inclusion of random variables as possible predictors generated different effects through the datasets. On the Micro 1 approach, as expected, the same pattern of sparsity was preserved, with the last two predictors included 1.6\% and 3.9\% of the times, respectively. On the Macro 1 and Micro 2 settings, despite the lack of a pattern of sparsity, the Spike-and-Slab performed reasonably regarding the inclusion of the random variables. They were included 12.2\% and 21.1\% of the times on the Macro 1, and 20.0\% and 18.7\% on the Micro2.
    
    On the other hand, the model performed poorly in identifying the irrelevance of the random variables on the Macro 2 (56.1\% and 55.2\%) and Finance 1 (71.0\% and 48.4\%) settings. It is interesting to notice, however, that on the Macro 2 the random variables ranked as 5\textsuperscript{th} and 6\textsuperscript{th} least included variables, from a total of 62. On the Finance 1, one of the random regressors was the least included among 18, while the other one ranked as the 3\textsuperscript{rd} more included.
    
    This experiment corroborates, at least for the Finance 1 and Macro 2 datasets, to the idea that the design of the model is itself inducing a high level of selection and shrinkage, not fulfilling the goal of allowing for shrinkage or sparsity in order to learn the best approach. Still, it is important to notice that in this subsection only one set of simulated variables was generated for each dataset, and that different results can be drawn depending on the generated predictors. However, the fact that two of the five settings presented a strong difference between the results and what would be expected suggest that similar outcomes would be achieved if the experiment was run more times, or with a different number of random variables. 
    
\begin{figure}[t]
        \centering
        \includegraphics[height=10cm, width=17cm]{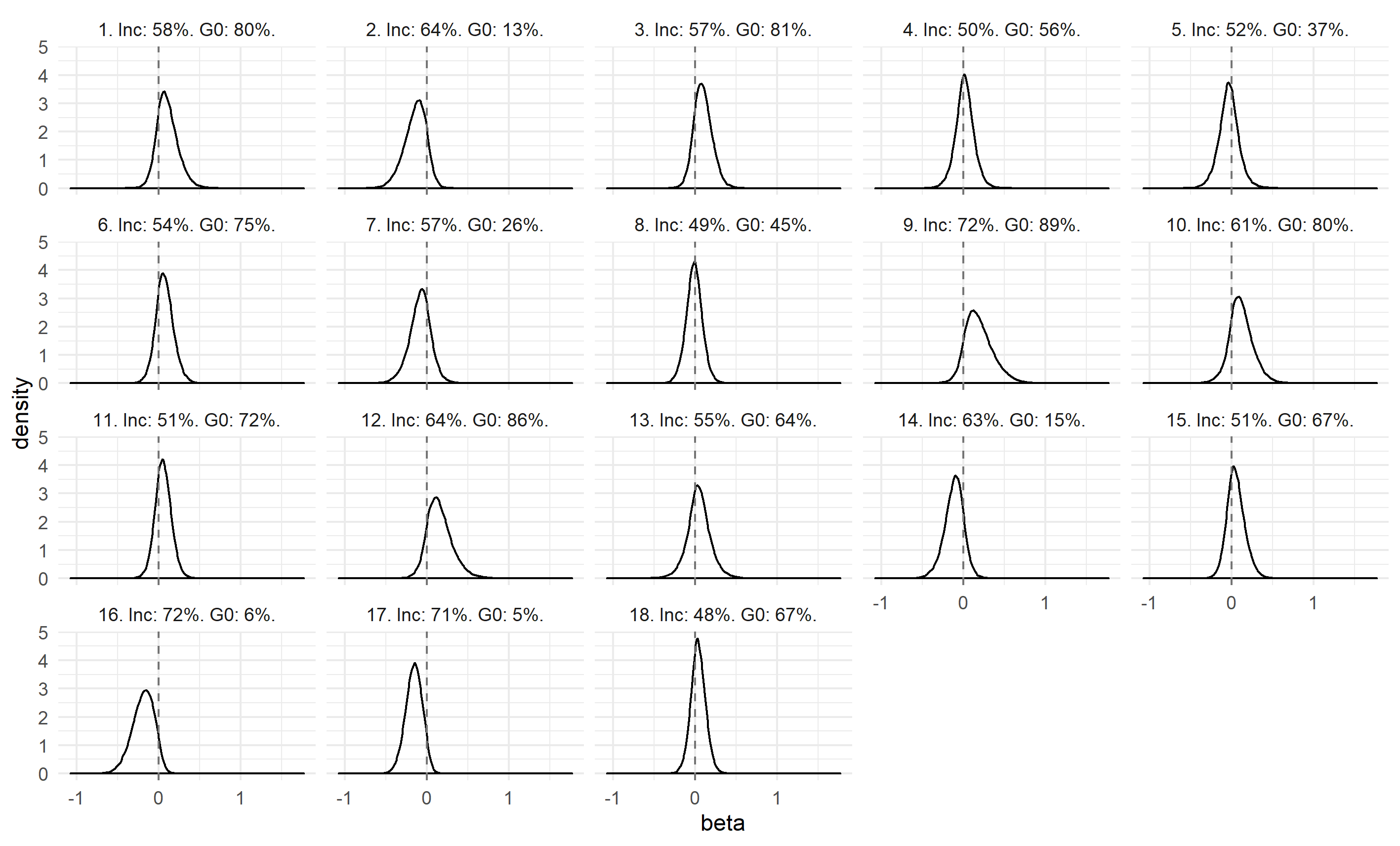}
        \caption{Posterior distribution of beta for the Finance 1 dataset with additional random variables. "Inc." means the probability of inclusion and "G0" the probability of being greater than zero.
Predictors 17 and 18 are randomly generated variables.
}
        \label{fig:rnd_beta}
    \end{figure}

    The graphics in figure \ref{fig:rnd_beta} bring the posterior distribution of $\beta_i | (\beta_i \neq 0)$, that is, the estimated density of the value of the coefficients $\beta$ for all the possible predictors, when included. Above each graphic is included the number of the regressor, the probability of inclusion in the model, and the probability of the coefficient to be greater than zero.
    
    It can be noticed that for the original predictors, from 1 to 16, the posterior distributions are extremely similar to the original graphics on figure \ref{fig:beta_fin1}, indicating that the inclusion of the new variables didn't interfere on the estimation of the other parameters. The posterior of predictor 18 is not surprising, given what was already discussed on section 3.1.5. Besides the probability of inclusion of 48\%, the distribution is concentrated on very small values of $\beta$, symmetric around zero, indicating that the likelihood of including the variable in the model with a very small coefficient is similar to the likelihood of excluding it. It once again corroborates with the thesis that the Spike-and-Slab incorrectly stimulates selection and shrinkage.
    
    Predictor 17, on the other hand, despite being completely random was included 71\% of times, 95\% of them with a negative value. Following the discussion on figure \ref{fig:beta_fin1}, it indicates that in fact a variable cannot be assumed as relevant just because of its degree of inclusion or lack of symmetry around zero. It interesting, however, to notice that the distribution is still very concentrated on small values of beta, with 90\% of the distribution being greater than $-0.29$. This is not the case for other predictors, such as 9, for which 25\% of the distribution of is greater than $0.29$, or predictor 12, for which 18\% of the distribution of beta is larger than $0.29$. That is, the distribution of these predictors, when included, is less concentrated around zero than the ones of the random variables.
    
    This experiment suggests, once again, that the design of the model itself is unable to clearly distinguish the possibilities of shrinkage or sparsity, possibly inducing the former, depending on the setting. Especially on the cases when the posterior distribution was closer to the prior distribution, that is, the model had a poor learning, on the Finance 1 and Macro 2 datasets, the model seems to have induced some shrinkage, what is made explicit by the high probability of inclusion of the randomly generated variables.
    
    \subsection{Fatter Tails: Using the t-student Distribution}
    
    One possible explanation for the results achieved on the last subsections can be related to the shape of the distribution of the coefficients of the predictors. By using a Gaussian distribution, the Spike-and-Slab prior could be inducing the posterior distribution of beta to be concentrated around zero, thus generating an ambiguity of whether the model should include or not a predictor, as both options end up being very similar if the distribution of $\beta_i | (\beta_i \neq 0)$ is concentrated for very small values of $\beta_i$.
    
    In their article, GLP recognize that a misspecification of the regression coefficients distribution can lead to a poor performance:
    
\begin{quote}
Our approach relaxes all sparsity and density constraints, and instead imposes some structure on the problem by making an assumption on the distribution of the non-zero coefficients. The key advantage of this strategy is that the share of non-zero coefficients is treated as unknown, and can be estimated. Another crucial benefit is that our Bayesian inferential procedure fully characterizes the uncertainty around our estimates, not only of the degree of sparsity, but also of the identity of the relevant predictors. The drawback of this approach, however, is that it might perform poorly if our parametric assumption is not a good approximation of the distribution of the non-zero coefficients. Even if we take this concern into consideration, at the very least our results show that there exist reasonable prior distributions of the non-zero regression coefficients that do not lead to sparse posteriors. 
    \cite{glp}
    \end{quote}
    
    To deal with the problem, they use simulated datasets to show that the model is capable of learning the degree of sparsity when using the Gaussian distribution for the regression coefficients, even on different settings for the data-generating process. Also, they explore the out-of-sample performance of their model, compared to sparse models. Although they reach interesting results on the simulated datasets, they do not explore the differences of changing the distribution of the non-zero coefficients on the real datasets.
    
    In order to further explore the question, we propose a change in the model, substituting the normal distribution in the Spike-and-Slab prior for a t-student distribution. A desirable feature of the t-student is that it presents fatter tails, that is, its density is higher for the values more distant to zero than in the normal distribution. Section 3.3.1 describes how the substitution was implemented, and the changes on the algorithm. Section 3.3.2 brings the results of the estimation, showing the posterior probability of inclusion of each predictor, and the posterior distribution of the coefficients once included.
    
    \subsubsection{Implementation}
    
    The substitution of the Gaussian for a t-student distribution is implemented by adding a latent variable $\lambda_i$ to the model. Specifically, we change the prior distribution of $\beta_i | \sigma^2, \gamma^2, q$, as described in section 2.1, for:
    
    $$\beta_i | \sigma^2, \gamma^2, \lambda_{i}^{2}, q \sim \left\{\begin{matrix}
    
    N(0,\sigma^2 \gamma^2 \lambda_{i}^{2}) & \text{with prob. }q \\
    
    0 & \text{with prob. } 1 - q 
    
    \end{matrix} \ \ \ i = 1, ..., k.\right.$$
    
    And set an Inverse-gamma prior distribution for $\lambda_{i}^{2}$:
    
    $$\lambda_{i}^{2} \sim IG\left( \frac{\nu}{2}, \frac{\nu}{2} \right)$$
    
    It can thus be shown that:
    
    $$\beta_i | \sigma^2, \gamma^2, q \sim \left\{\begin{matrix}
    
    t_\nu(0,\sigma^2 \gamma^2) & \text{with prob. }q \\
    
    0 & \text{with prob. } 1 - q 
    
    \end{matrix} \ \ \ i = 1, ..., k.\right.$$
    
    Where
    
    $$Var[\beta_i] = \frac{\nu}{\nu - 2} \sigma^2 \gamma^2$$
    
    Instead of learning the parameter $\nu$, we estimate the model for the pre-defined values of 4, 10, 30, 100 and 500. Given the shape of the t-student distribution, the prior distribution of $\beta_i | \sigma^2, \gamma^2, q$ thus has very fat tails for $\nu = 4$, and a very similar shape to the normal distribution when $\nu = 500$.
    
    The estimation algorithm has few changes. Taking as a basis the algorithm developed in Appendix A of GLP, and preserving the same notation, $\overline{v}_x$ is redefined as:
    
    $$\overline{v}_x \equiv E[\sigma_{i, i}] \frac{\nu}{\nu - 2}$$
    
    This way, considering the redefinition of $\overline{v}_x$, the conditional posterior distributions of $R^2$ and $q$, $\phi$, $z$ and $\sigma^2$ are all unchanged. The conditional distribution of $\beta$ is now induced by:
    
    $$\frac{\beta_i}{\sqrt{\lambda^{2}_i}} =: \beta^{*}_i | Y, \phi, \sigma^2, R^2, q, z \sim \left\{\begin{matrix}
    
    t_\nu(0,\sigma^2 \gamma^2) & \text{with prob. }q \\
    
    0 & \text{with prob. } 1 - q 
    
    \end{matrix} \ \ \ i = 1, ..., k.\right.$$
    
    Finally, the conditional distribution of $\lambda^2_i$ is given by:
    
    $$\lambda_i^2 | \nu, \beta_i, \sigma^2, R^2 \sim IG\left( \frac{\nu + 1}{2}, \frac{\nu + \beta^2_i / \sigma^2 \gamma^2}{2} \right)$$
    
    \subsubsection{Results}
    
    \begin{figure}[t]
    \centering
        \includegraphics[width=16cm, height=15cm]{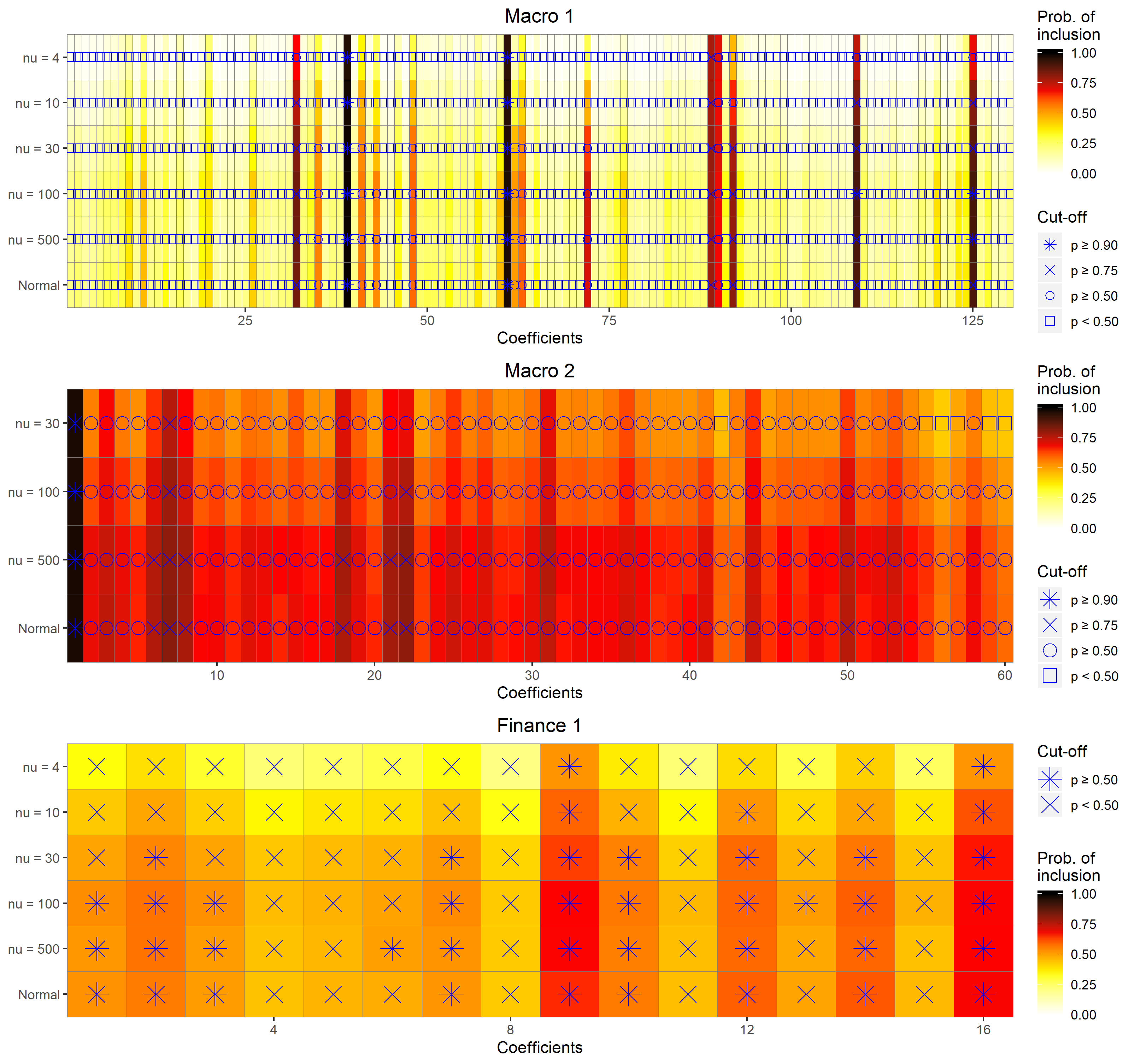}
        \caption{Probability of inclusion of each predictor for the Macro 1 and 2 and Finance 1 datasets. Each columns is a predictor and each row one model, varying the number of degrees of freedom $\nu$.}
        \label{fig:t_vars1}
    \end{figure}
    
    \begin{figure}[t]
    \centering
        \includegraphics[width=16cm, height=10cm]{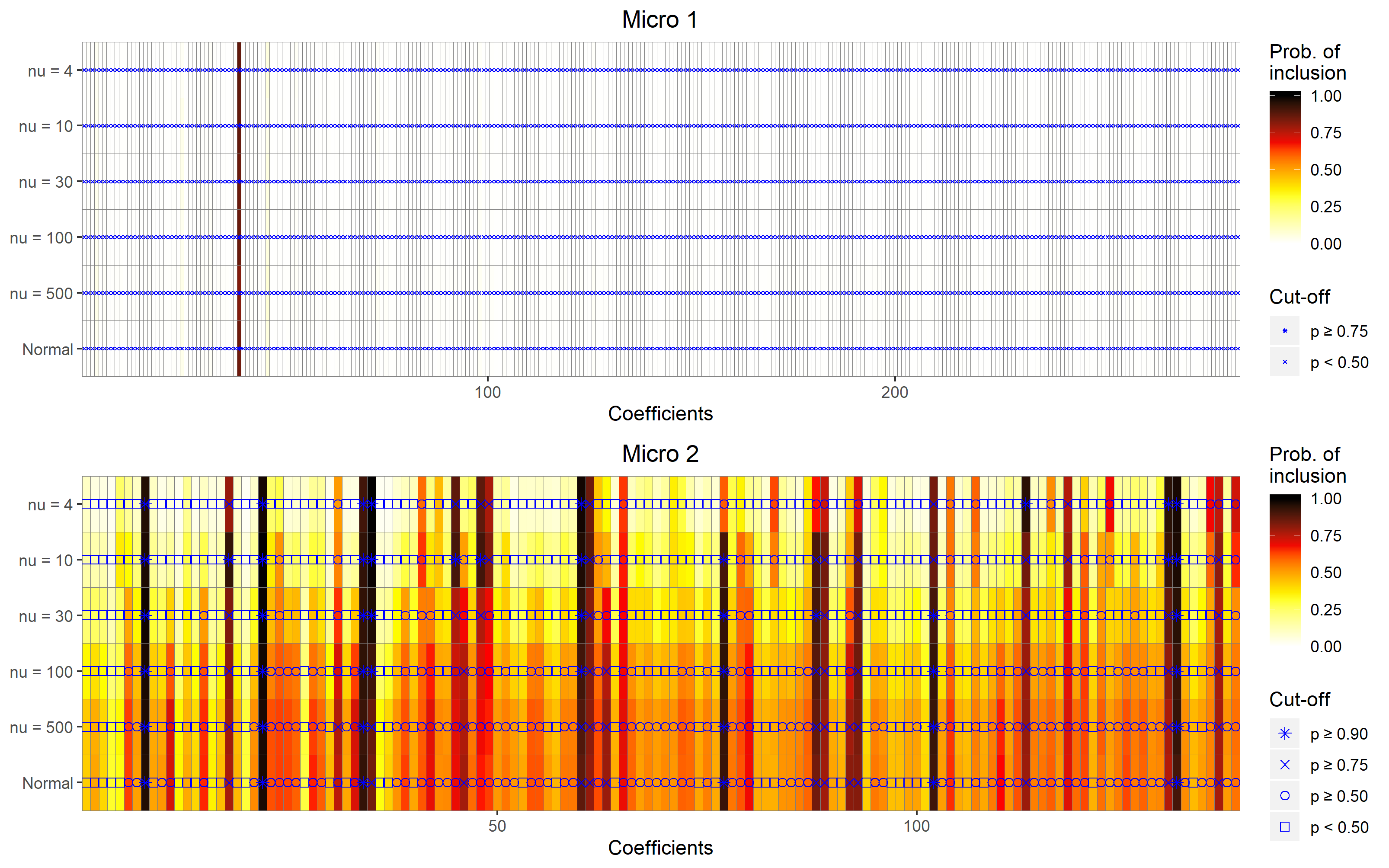}
        \caption{Probability of inclusion of each predictor for the Micro 1 and 2 datasets. Each columns is a predictor and each row one model, varying the number of degrees of freedom $\nu$.}
        \label{fig:t_vars2}
    \end{figure}
    
    Figures \ref{fig:t_vars1} and \ref{fig:t_vars2} bring the estimated probability of inclusion of each regressor for all the five datasets, considering different values for the number of degrees of freedom of the prior t-student distribution $\nu$. 
    
    The graphic is an adaptation of the already used in figures \ref{fig:vars} and \ref{fig:rnd_vars}, but now instead of using a whole stripe for each coefficient, the stripes are divided in rectangles, with each row representing a different value for $\nu$, while as usual each column represents one possible predictor. The color of the heatmap is unchanged, using light colors for a small probability of inclusion, and a darker tone as the probability increases. 
    
    It was also included a "cut-off" indicator, to help interpreting the dimension of the probability of inclusion. The cutting levels are of 50\%, 75\% and 90\%. Therefore, for example, on the Finance 1 dataset in figure \ref{fig:t_vars1}, the first row, corresponding to the heavy-tailed t-student with only 4 degrees of freedom, contains only two coefficients included more than 50\% of times, and none included more than 75\%. In the Macro 2 setting in the same figure, in the first row only the first predictor is included with a probability higher than 90\%, and the probability of inclusion of the seventh predictor is between 75\% and 90\%, while for all others it is smaller than 75\%. On the last row, the Normal distribution, seven predictors are included between 75\% and 90\% of the times.
    
    As expected, the last two rows are virtually equal for all the settings, reflecting that a t-student distribution with $\nu = 500$ can be approximated by a normal distribution. Slight variations between them are due to the limited size of the drawn MCMC. Also, it is not surprising that as the number of degrees of freedom decrease, the average probability of inclusion also decreases. It reflects the fact that the distribution's tails are heavier for small values of $\nu$, and so there's a greater distinction between including or not a regressor, as the likelihood that the coefficient be around zero is relatively smaller. By itself, this result again endorses the suspicion that the Spike-and-Slab, as originally defined, induces selection and shrinkage.
    
    Still, it is interesting to notice that, in some cases, the use of the t-student doesn't seem to have changed the pattern of variable selection, but only reduced the overall probability of inclusion. This seems to be the case of the Finance 1 dataset, in figure \ref{fig:t_vars1}, for which the probability of inclusion was very similar to the values of $\nu$ from 30 to 500, and for the Macro 2 dataset in the same figure, for which the pattern of the most included variables seems unchanged through the rows.
    
    The result for the Micro 1 dataset in figure \ref{fig:t_vars2} is also not surprising. Since the normal distribution was enough to identify the dominance of one single variable over the others for fitting the model, it was expected that the more restrictive t-student distribution wouldn't allow for selection of more variables, or block the selection of the single dominant predictor.
    
    Finally, the Macro 1 and Micro 2 settings show an interesting behavior. In Macro 1 in figure \ref{fig:t_vars1}, while most of the variables have a probability of inclusion smaller than 50\% even for the normal distribution case, some variables that are included with a high frequency in the last row are excluded most of the times on the first rows. Moreover, it happens without changing the probability of inclusion of other variables, that is, there is a change in the pattern of variable selection. If, say, an economist was to believe in the selection power of the model with a t-student distribution with 4 degrees of freedom, he would find that only 7 of 130 available predictors are relevant - that is, included more than 50\% of times -, what could be interpreted as a sparse model.
    
    A similar although weaker effect can be seen in the Micro 2 dataset in figure \ref{fig:t_vars2}. While the normal distribution setting shows no clear pattern of variable selection, the t-student cases are capable of more clearly selecting some predictors, decreasing the probability of selection of several variables while preserving a high probability for others, that is, it changes the pattern of variable selection. So, for example, while for the normal distribution 106 of 138 predictors are selected more than 50\% of the times, for the case when $\nu = 4$ only 30 are selected, and 34 for $\nu = 10$.
    
    Although these results are insufficient to conclude that the Spike-and-Slab with a t-student distribution can be used to identify whether sparsity or shrinkage should be chosen for a dataset, they are a strong evidence that the use of the normal distribution is insufficient to draw such conclusion, as the use of this prior distribution induces high levels of variable selection with shrinkage.
    
\begin{figure}[t]
        \centering
        \includegraphics[height=10cm, width=17cm]{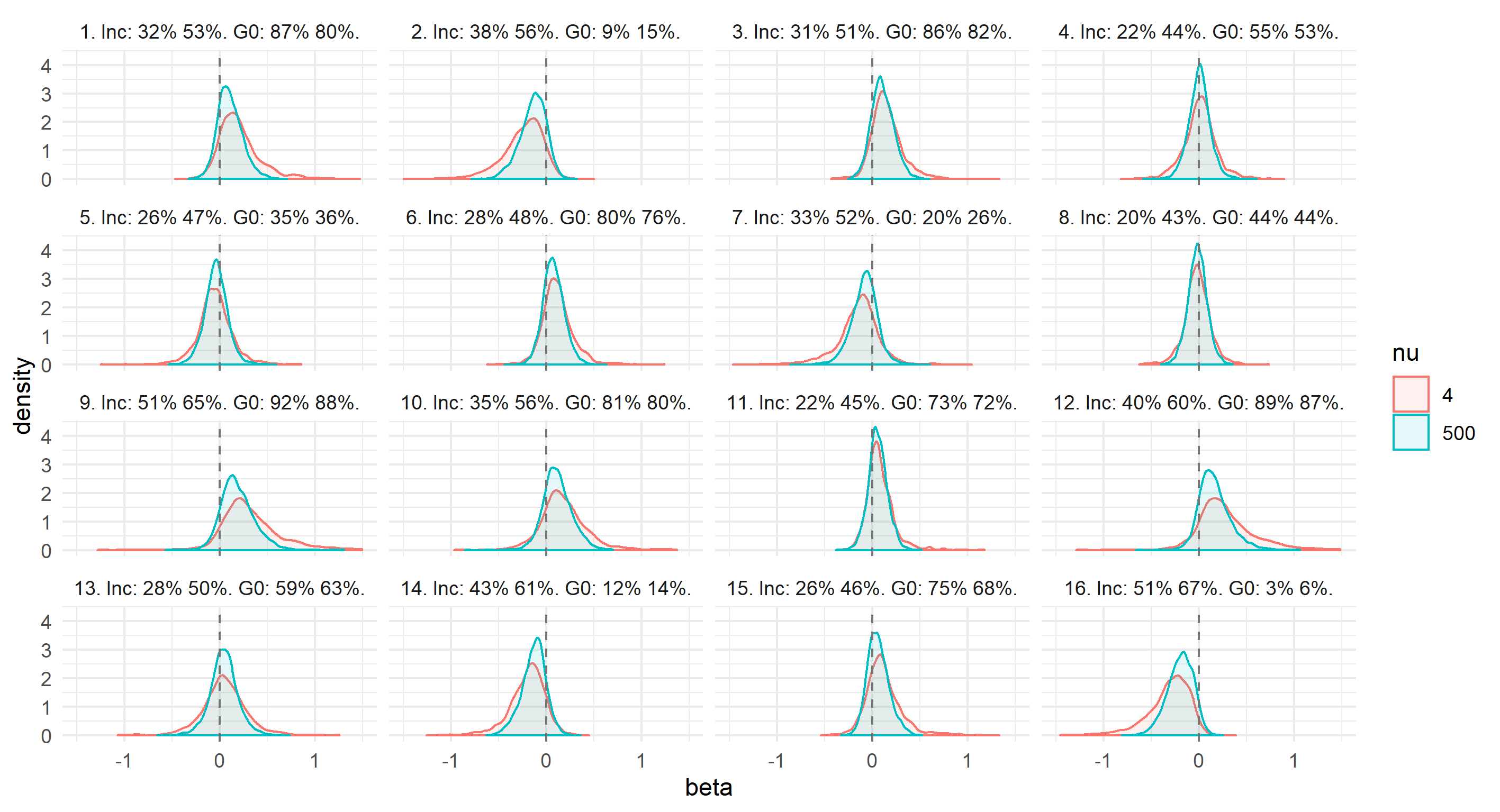}
        \caption{Posterior distribution of beta for the Finance 1 dataset for the t-student prior distribution. "Inc." means the probability of inclusion and "G0" the probability of being greater than zero. The first value is for the case $\nu = 4$, and the second for $\nu = 500$.}
        \label{fig:t_beta}
    \end{figure}

    Additional to the variable selection pattern, figure \ref{fig:t_beta} updates figure \ref{fig:beta_fin1}, for the Finance 1 dataset, with the t-student as the prior distribution. It compares the posterior distribution of the coefficients beta for each predictor once they are included. The title of each graphic brings first the index of the variable - the same used in figures \ref{fig:t_vars1} and \ref{fig:t_vars2} -, the probability of inclusion "Inc." - the same from the last figures - and  "G0", the percentage of the distribution concentrated in positive values, respectively to the case when $\nu = 4$ and $\nu = 500$ (the approximation to the normal distribution).
    
    The figure reveals that the probability of inclusion of all the variables decrease significantly, and the distribution with $\nu = 4$ becomes more asymmetric and skewed for all of the predictors. Concerning the selection problem discussed in section 3.1.5, the t-student by itself, even in the extreme case of only four degrees of freedom, still doesn't seem enough to solve the ambiguity in the model for choosing whether a predictor should be included or not. It happens because even when a variable is included in the model, its estimated value if very close to zero has almost the same impact in the model, thus resulting in a similar likelihood for both inclusion in the model with a very small coefficient or exclusion. This effect appears to be overestimating the probability of inclusion of the regressors, thus resulting in a difficulty in identifying the presence of sparsity, that is, the set prior distributions seem to be inducing density and shrinkage, and underestimating the possibility of sparsity.
    
    \subsection{A Simulation Study}
    
    Based on the results from section 3.2, which showed a poor performance of the model in excluding completely randomly generated predictors for some of the datasets, and the new model proposed in section 3.3, this section proposes a simulation study. We simulate a dataset with the same dimensions of the Finance 1 setting, with 68 observations and 16 covariates. We predefine the value of the coefficient beta for the first three predictors, and set the other 13 to be exactly equal to zero. Therefore, the model would perform accurately if correctly included only the first three regressors.
    
    The data generating process is as follows. We first draw a random vector $\varepsilon$ from a normal distribution and set the values for $\beta_1$, $\beta_2$ and $\beta_3$. Moreover, we calculate the response variable as the sum of the first three covariates multiplied by their respective coefficients plus an error term, and consider six scenarios, varying the variance of the error term, $\sigma^2_{\varepsilon}$. That is, given that $X$ is the dataset, where $X_{i, j}$ is the value of predictor $j$ for individual $i$, draw:
    $$X_{i, j} \sim N(0, 1)$$
    $$\varepsilon_i^* \sim N(0, 1)$$
    
    Set:
    $$\beta_1 = -0.86,\ \beta_2 =  0.64,\ \beta_3 =  0.89$$
    
    Calculate:
    $$y_i^{(s)} = \beta_1 X_{i, 1} + \beta_2 X_{i, 2} + \beta_3 X_{i, 3} + \sigma_{\varepsilon}^{(s)} \varepsilon_i^*$$
    
    For $s \in \{1, ..., 6\}$ and $i \in \{1, ..., 68\}$. We define:
    
    $$\sigma_{\varepsilon}^{(s)} = 0.75 s$$
    
    That is, for example, $\sigma_{\varepsilon}^{(1)} = 0.75$, $\sigma_{\varepsilon}^{(3)} = 2.25$, and so on. This number represents the uncertainty on the dataset: if $\sigma_{\varepsilon}$ is very small, any model that allows for variable selection should perform reasonably in selecting only the three first predictors; if $\sigma_{\varepsilon}$ is large, different models should perform differently in selecting the appropriate variables, some better than others. After $y$ and $X$ are defined, they are all scaled to have exactly zero mean and standard deviation one, repeating the same approach developed in GLP and in the previous sections.
    
    \begin{figure}[t]
    \centering
        \includegraphics{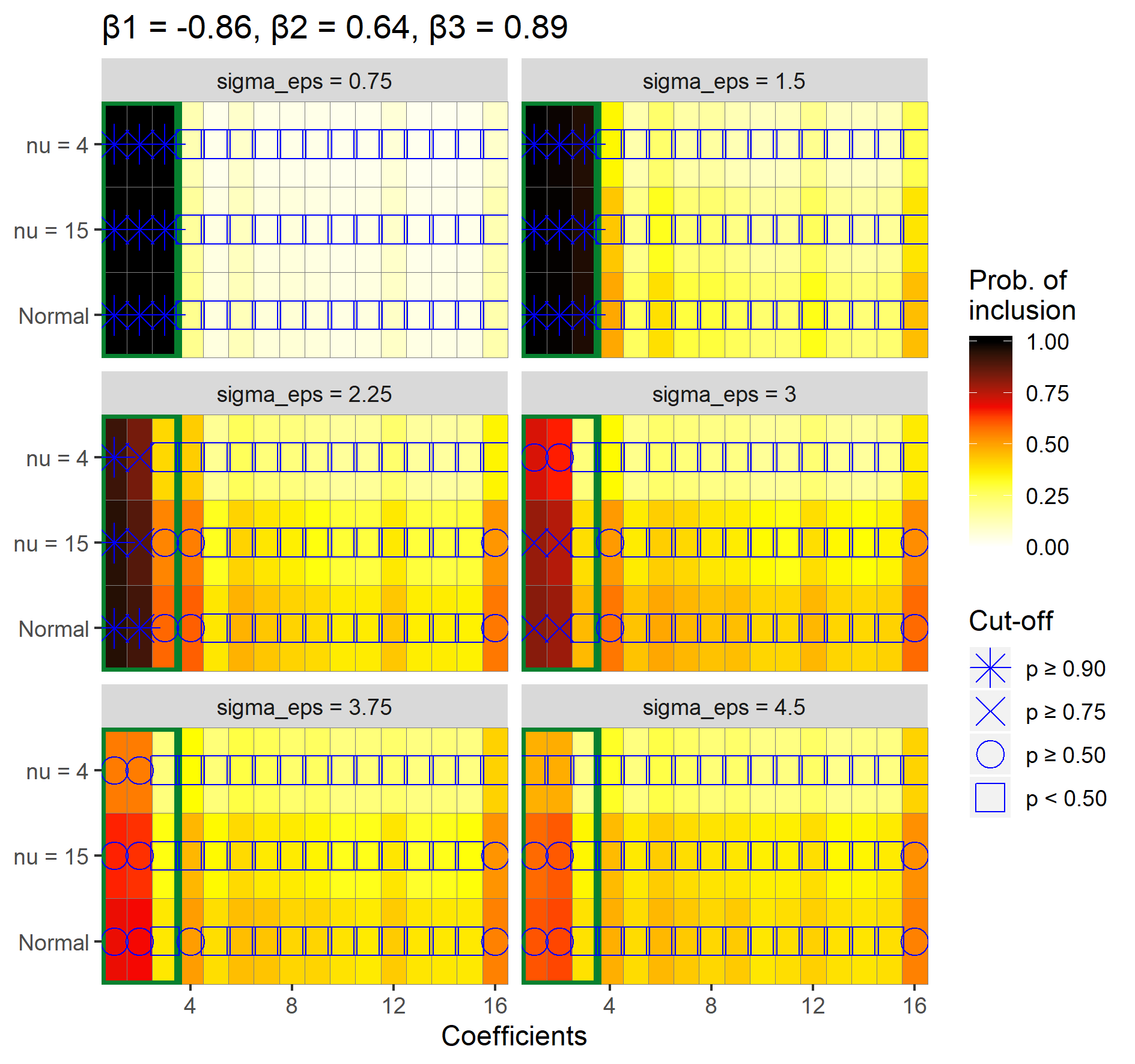}
        \caption{Probability of inclusion of each predictor for simulated datasets. Each columns is a predictor and each row one model, varying the number of degrees of freedom $\nu$. Each block represents one dataset, varying the standard error of the error term, $\sigma_{\varepsilon}$.}
        \label{fig:sim_vars}
    \end{figure}
    
    \begin{figure}[t]
        \centering
        \includegraphics[height=10cm, width=17cm]{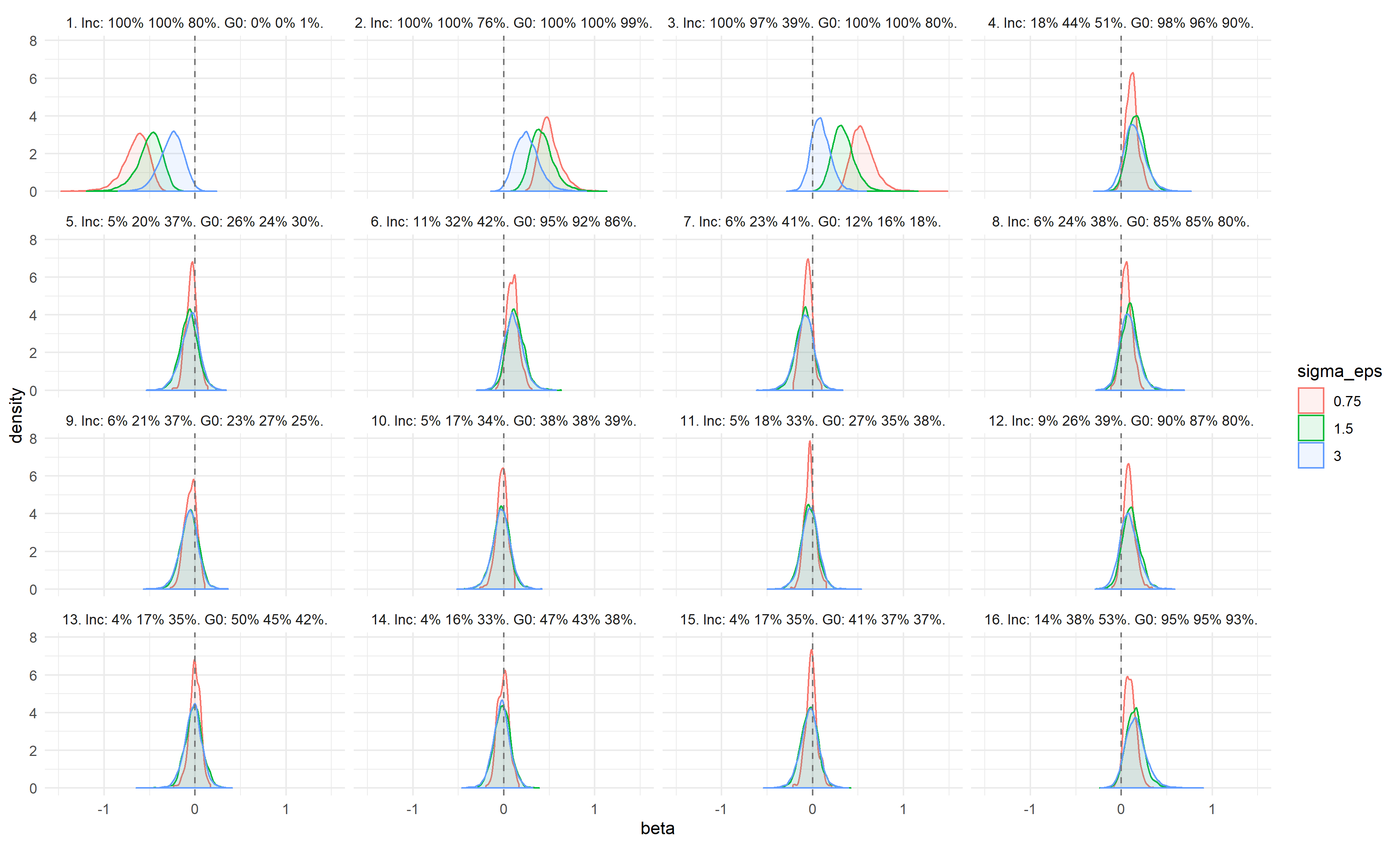}
        \caption{Posterior distribution of beta for three simulated datasets with Gaussian regression coefficients, varying $\sigma_{\varepsilon}$. "Inc." means the probability of inclusion and "G0" the probability of being greater than zero. The order of the values follow the order of sigma\_eps.}
        \label{fig:sim_beta}

    \end{figure}
    
    The graphics in figure \ref{fig:sim_vars} follow the same approach as the one in figures \ref{fig:t_vars1} and \ref{fig:t_vars2}, described in section 3.3.2. As expected, when the variance of the error term is small, such as in the case where $\sigma_{\varepsilon} = 0.75$, the model easily selects the three truly relevant predictors with almost 100\% of probability, and all others with virtually zero. As this variance increases, when $\sigma_{\varepsilon} = 1.50$, the model is still accurate in selecting correctly only the relevant variables most of the times, but it can also be seen that the other predictors are included with higher probability, but that never reach 50\%.
    
    The model starts failing for $\sigma_{\varepsilon} = 2.25$, when all three models fail in selecting the third predictor more than 75\% of times. The predictors of number 4 and 16 are also incorrectly selected more than 50\% of times for the normal and $\nu = 15$ settings, what doesn't happen for the $\nu = 4$ case. On the other hand, this more restrictive model also fails more than the other in selecting the third variable, what happens less than 50\% of times. A similar effect happens as the standard deviation of the error term increases. For $\sigma_{\varepsilon} = 4.5$, for example, the three settings fail in the selection, and for $\nu = 15$ and the normal distribution, predictor 16 becomes almost as important as predictors 1 and 2, while predictor 3 is rarely selected. For $\nu = 4$, all the variables are selected with a very low rate, with predictors 1, 2 and 16 being the most selected.
    
    These results might indicate that the model has limited capacity in distinguishing patterns of sparsity, or at least that the datasets considered might have an elevated level of uncertainty (portion of the response variable explained by unobservables), such that few can be learned from the use of econometric models. Even though making effort to make statistical learning even in difficult settings is a major task of statisticians and economists, it is worth mentioning that an extreme scenario can be the case in at least some of the datasets considered, as they are very "small" considering the number of observations, and "big" if considered the large number of possible predictors.
    
    The graphics in figure \ref{fig:sim_beta} bring the posterior density distribution of the coefficients beta for the 16 simulated predictors, in the same approach as figure \ref{fig:t_beta}. It is interesting to notice that the distributions for predictors 1 to 3, for the cases $\sigma_{\varepsilon} = 0.75$ and $\sigma_{\varepsilon} = 1.50$, are very offset from zero, correctly identifying the true predefined parameters. It is interesting to notice, though, that as uncertainty grows, also grows shrinkage, and all the distributions converge towards zero. This is especially the case for regressor 3, whose distribution concentrates around zero, what also leads to a great drop in the probability of inclusion in the model. This result once again corroborates with the thesis that the model creates an ambiguity between inclusion with shrinkage or exclusion, as the likelihood of both becomes very similar, and the model fails in learning the correct approach.
    
    As for the other regressors, we notice that the concentration of the posteriors in one of the sides of zero is uninformative. For example, predictor 12, when included, has more than 80\% of its distribution concentrated only in positive values of beta for all of the three settings evaluated. Still, no inference can be made about this feature of the shape of the distribution, as the coefficient $\beta_{12}$ is known to be exactly zero. It also collaborates with the hypothesis previously presented that the fact that the distribution is very concentrated on small values of beta, close to zero, is a more important indicator than how offset from zero the distribution is, showing the unimportance of a predictor for the model. The fact that the distributions for virtually all of the unimportant variables are very concentrated around zero implies, again, that the probability of inclusion is probably overestimated due to the similar likelihood of excluding the predictor or including it with a very small coefficient (high degree of shrinkage).
    
    \section{Conclusion}
    
    This paper reassessed the model proposed in GLP, through a more detailed look into the posterior distributions, and the proposition of three experiments. First, after adding random variables to the datasets and reevaluating the model, it was found that in some of the settings the model was unable to distinguish a completely random regressor from the other available economic series, even privileging a random predictor in one of the settings in spite of others. 
    
    Second, a modification was proposed to the model, substituting the prior Gaussian distribution of the coefficients of the model for a t-student distribution. It was shown that, depending on the the number of degrees of freedom of the distribution, a sparse model is naturally distinguished among the predictors for one of the datasets, unlike the result obtained with the normal distribution. For the other datasets, the effect was not homogeneous, but the model with the t-student showed overall an improvement on the pattern of what variables should be excluded. Finally, the simulation study indicates that the Spike-and-Slab is biased towards selecting more predictors and shrinking their coefficients.
    
    All the experiments corroborate with the idea that the model is, itself, inducing variable selection and shrinkage. The mechanism through which it happens could be that the likelihood of excluding an irrelevant predictor, or including it with a very small coefficient, is very similar. Thus, the evidences suggest that the Spike-and-Slab prior distribution is itself inducing density, and thus does not fulfill the objective of GLP of evaluating whether density or sparsity is more adequate for a given dataset.
    
    It is important to notice that this paper does not contradict the conclusion achieved by GLP, but brings evidence to show that the model proposed is not robust to find the conclusion that economic datasets are not informative enough to identify a conclusive pattern of sparsity among many possible predictors. It was indeed shown that an unique set of a few relevant predictors was identified for the Micro 1 dataset, for all approaches considered, and also for the Finance 1 dataset, if a heavy-tailed t-student is used in the prior distribution of the coefficients. For other cases, such as the Micro 2 dataset, the t-student also had a much better performance in excluding irrelevant predictors. Still, our findings corroborate with GLP conclusion that, without statistical evidence, sparsity should not be simply assumed in an economics dataset.
    
    But even more than that, the evidence brought by this paper shows that the setting of the prior distribution can drastically improve the performance of the model in detecting sparsity, and it thus indicates that further methods can help answer the question of whether sparsity can be used to model a given economics dataset or not.
    
    Finally, we conclude that the use of the Spike-and-Slab prior, such as proposed, is misleading if the goal is to find evidence of sparsity in an economics dataset. The model, by inducing shrinkage and selection, incorrectly provokes an illusion that sparsity is nonexistent: the illusion of the illusion of sparsity.
    
\bibliography{refs}
\bibliographystyle{apalike}

\end{document}